**Ferroelectric hysteresis in singly aligned graphene-hBN moiré superlattices**

*Bao Q. Tu, * Tanweer Ahmed, * Garen Avedissian, Suzanne Lancaster, Mayank Sharma, Kenji Watanabe, Takashi Taniguchi, Fèlix Casanova, Marco Gobbi and Luis E. Hueso**

Bao Q. Tu, Tanweer Ahmed, Garen Avedissian, Suzanne Lancaster, Mayank Sharma, Fèlix Casanova, Luis E. Hueso,
CIC nanoGUNE,
BRTA,
20018 Donostia-San Sebastian, Basque Country, Spain.
E-mail: qb.tu@nanogune.eu, t.ahmed@nanogune.eu, l.hueso@nanogune.eu

Bao Q. Tu, Mayank Sharma
Departamento de Polímeros y Materiales Avanzados: Física, Química y Tecnología,
University of the Basque Country (UPV/EHU),
Paseo Manuel de Lardizábal 3, 20018, Donostia-San Sebastian, Spain.
E-mail: qb.tu@nanogune.eu

Kenji Watanabe,
Research Center for Electronic and Optical Materials,
National Institute for Materials Science, 1-1 Namiki,
Tsukuba 305-0044, Japan.

Takashi Taniguchi,
Research Center for Materials Nanoarchitectonics,
National Institute for Materials Science, 1-1 Namiki,
Tsukuba 305-0044, Japan.

Marco Gobbi,
Centro de Física de Materiales (CFM-MPC)
Centro Mixto CSIC-UPV/EHU,
20018 Donostia-San Sebastian, Basque Country, Spain.

Marco Gobbi, Fèlix Casanova, Luis E. Hueso,

IKERBASQUE,
Basque Foundation for Science, 48009 Bilbao, Basque Country, Spain.
E-mail: l.hueso@nanogune.eu

**Abstract.**
Ferroelectric materials have the unique ability to maintain an electric polarization which can be reversed under an external applied electric field. This property makes them valuable for applications such as non-volatile random-access memories, transducers, actuators and electro-optic modulators. Recently, emergent unconventional ferroelectricity has been demonstrated in moiré superlattices of bilayer graphene and hexagonal boron nitride (hBN) hosting non-centrosymmetric stacking order. Whether this phenomenon is also present in non-centrosymmetric single layer graphene (SLG)-hBN moiré superlattices is still under debate. Here we demonstrate a ferroelectric response in an SLG-hBN moiré superlattice. Through Hall measurements, we pinpoint the origin of the hysteretic behavior to abnormal charge screening due to the moiré superlattice band and estimate the spontaneous polarization magnitude in the moiré superlattice structure. Temperature dependent measurements confirm that the hysteretic behavior persists from 2K up to room temperature, opening opportunities for high-mobility, ultrathin non-volatile devices.

## 1. Introduction

Ferroelectric materials, defined by their spontaneous and switchable electric polarization under an external electric field, are foundational to a wide array of technologies. Their bistable polarization states enable vast applications in non-volatile memories[1,2], field-effect transistors (FET)[3,4], solar cells, sensors or photonics devices[5–7]. The discovery of ferroelectricity in Rochelle salt in 1921[8], followed by observations in $KH_2PO_4$ by Busch and Scherrer in 1935[9], initiated a century-long exploration of ferroelectric phenomena. Since then, numerous ferroelectric materials have been identified, including classic 3D perovskites such as $BaTiO_3$[10], PZT[11] and $BiFeO_3$[2], organic materials such as PVDF[12], and oxides such as $HfO_2$ and $ZrO_2$[13,14]. Recently, the emergence of 2D van der Waals (vdW) ferroelectric materials has introduced a new paradigm in the field.[15–22] The layered structure of these ferroelectric vdW materials allows them to be exfoliated down to a monolayer or a few layers, providing an excellent platform to study ferroelectricity in low-dimensional systems. Furthermore, the ability to restack 2D layers into vdW heterostructures adds another tuning knob for the observation of novel mechanisms and unique properties.[21,23–26]

One particularly transformative direction in vdW ferroelectrics involves the construction of moiré superlattices, formed when two atomically thin layers are stacked with a small twist angle or lattice mismatch.[27–30] These moiré superlattices have recently been demonstrated to give rise to interfacial ferroelectricity, a new avenue for controlling charge polarization at the nanoscale.[31–33] This distinct purely electronic ferroelectric effect has been experimentally observed in graphene-based systems, including quadrilayer graphene[34], bilayer graphene (BLG)[32], and more recently, even in single-layer graphene (SLG)[31], when both the top and bottom interfaces are precisely aligned (doubly aligned) with hexagonal boron nitride (hBN). The observed polarization switching is believed to originate from interlayer charge redistribution and layer-specific charge polarization, driven by strong electron–electron interactions in the moiré-engineered band structure. However, whether such ferroelectricity can exist in a singly-aligned SLG/hBN moiré system is still unclear.

Here we demonstrate that electronic ferroelectric hysteresis does indeed emerge in SLG/hBN with only one moiré superlattice interface. By disentangling the effects of interfacial charge polarization and moiré potential modulation through Hall measurements, we demonstrate that the hysteresis peaks arise from crossings of the charge neutrality point (CNP), allowing us to estimate the polarization strength associated with moiré-induced ferroelectricity. Our results exhibit a counterclockwise hysteresis loop, characteristic of conventional ferroelectric behavior. This ferroelectric effect persists over a wide temperature range (2K to 300K), highlighting its robustness and potential for applications at room temperature.

## 2. Results and discussion

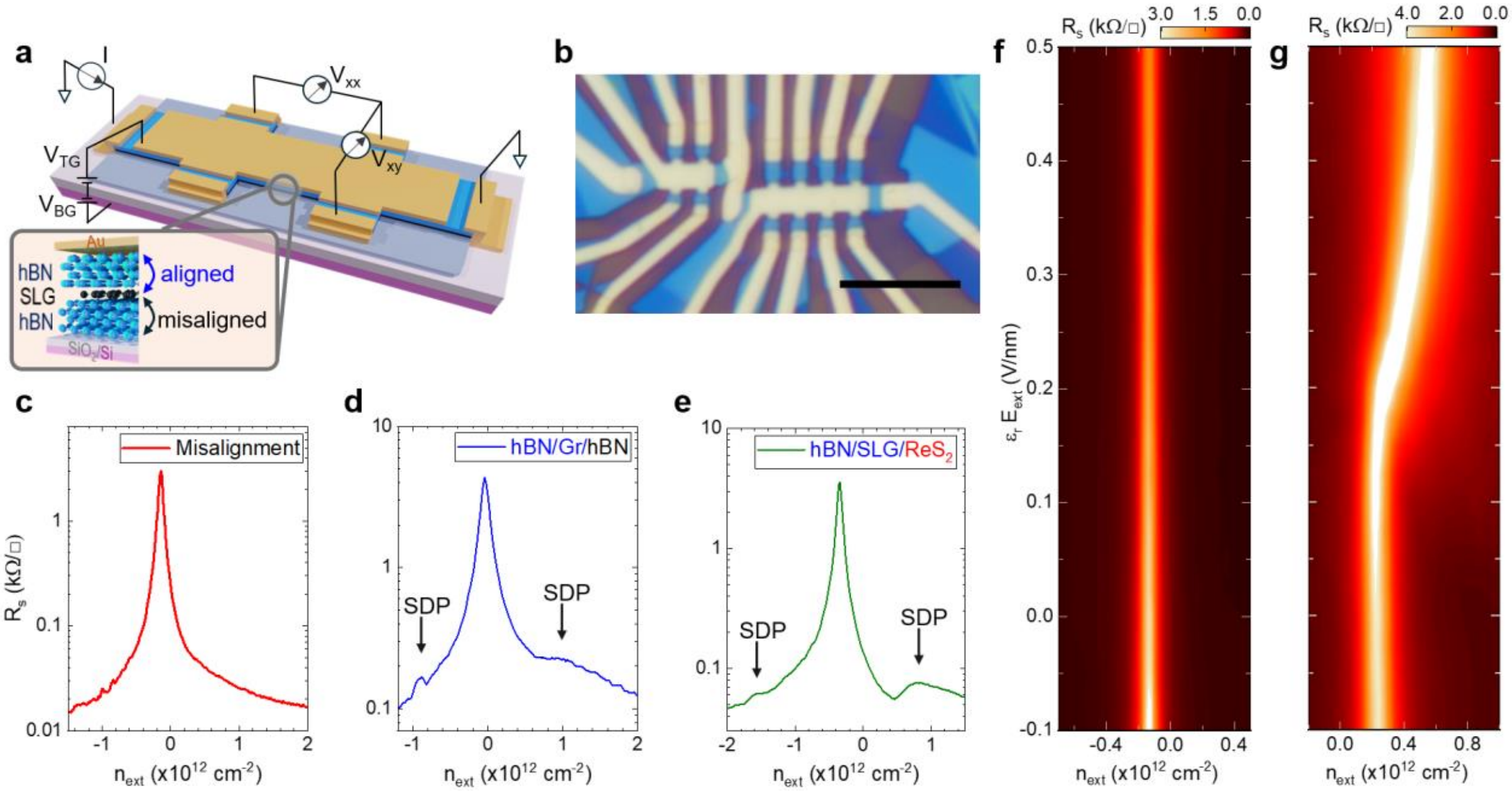

**Figure 1.** **(a)** Device schematic and measurement configurations. The inset shows the cross-section of device B1. (**b)** Optical microscope image of devices B1 and B2 with a 10 μm scalebar. **c,d,e**, Transfer characteristic of three different devices: **(c)** Device A: misaligned top and bottom hBN/SLG interfaces; **(b)** Device B1: singly-aligned hBN/SLG/hBN; **(d)** Device B2: singly-aligned hBN/SLG/$ReS_2$. **(e)** $n_{\mathrm{ext}} - E_{ext}$ phase space of $R_{\mathrm{s}}$ of Device A. **(f)** $n_{\mathrm{ext}} - E_{ext}$ phase space of $R_{\mathrm{s}}$ of Device B1.

All heterostructures were fabricated using the dry transfer pickup-and-drop technique.[35] The single-layer nature of graphene was confirmed by Raman spectroscopy and quantum Hall measurements (see SI 2). In Device A, an encapsulated SLG/hBN was fabricated with misaligned top and bottom interfaces (no moiré) as a reference. Device B1 and B2 have aligned top interfaces and misaligned bottom interfaces. Devices B1 and B2 were constructed with an aligned top hBN/SLG interface and a misaligned bottom interface to induce a moiré pattern only at the top interface. To further suppress any potential moiré superlattice formation at the bottom hBN/SLG interface, a few-layer-thick triclinic rhenium disulfide ($ReS_2$)[36,37] was intentionally inserted between the SLG and the bottom hBN, acting as a spacer to effectively disrupt any periodic potential from the bottom layer. In addition, because $ReS_2$ exhibits much higher resistivity than graphene, the transport signal originates predominantly from the SLG channel. Detailed fabrication procedures are provided in Supplementary Information (SI) section 1. **Figure 1**a schematically depicts a typical dual-gated device with the measurement circuit diagram. **Figure 1**b presents an optical micrograph of the devices B1 and B2. By controlling both top and bottom gates simultaneously, we independently control the external carrier concentration ($n_{\text{ext}}$) in the channel and the vertical external electric field ($E_{ext}$)[33,38], see **Methods** section for more details. The single layer thickness of graphene was confirmed by a combination of Raman spectroscopy and quantum Hall measurements (see SI section 3). The variation of sheet resistance ($R_{\text{S}}$) *vs.* $n_{\text{ext}}$ at $\varepsilon_r E_{ext} \approx 0$ from samples A and B1 are presented in **Figure 1**c and d, respectively. The typical $R_{\text{S}}$ vs. $n_{\text{ext}}$ is observed in Device A, with a maximum at the charge neutrality point (CNP). In Device B1, $R_{\text{S}}$ vs. $n_{\text{ext}}$ hosts additional peaks on both sides of the CNP, which are identified as secondary Dirac points (SDP), appearing due to the formation of moiré superlattice.[39,40] SDPs are also present in the aligned hBN/Gr/$ReS_2$/hBN sample **Figure 1**e. Based on the relative positions of SDP from CNP, we can calculate the moiré wavelength $\lambda = 21.2 \pm 0.3$ nm for device B1 and $\lambda = 19.4 \pm 0.3$ nm for device B2. The moiré wavelength extracted from the carrier concentration at the satellite Dirac points slightly exceeds the theoretically predicted value at perfect alignment ($\theta = 0$). This variation may arise from a combination of strain and experimental variations in the lattice constant of hBN in our sample. Further details are provided in the Supporting Information SI 4.

In **Figure 1**f, we demonstrate the variation of the sheet resistance $R_{\text{S}}$ as a function of $n_{\text{ext}}$ and $D$ ($n_{\text{ext}} - E_{ext}$ phase space) for Device A. We do not observe any significant variation of the CNP as a function of $E_{ext}$. The $n_{\text{ext}} - E_{ext}$ phase space of $R_{\text{S}}$ from Device B1 (singly aligned SLG/hBN) is presented in **Figure 1**g, which is strikingly different from that of Device A (misaligned SLG/hBN) in **Figure 1**f. Here, the CNP deviates at $\varepsilon_r E_{ext} = 0.2$ (V/nm) shifting from $n_{CNP} = 0.35\ (\times 10^{12}\ \text{cm}^{-2})$ to $n_{CNP} = 0.55\ (\times 10^{12}\ \text{cm}^{-2})$. This CNP deviation indicates a preferential screening of the electric field of one of the gates, in this case, the top gate. Such screening and shift in the CNP have been previously reported in doubly-aligned SLG/hBN and BLG/hBN moiré ferroelectric devices. The focus of this work is to understand whether such an effect in our singly-aligned SLG/hBN heterostructure is also related to the emergence of interfacial ferroelectricity. In SI section 4, we present $n_{\text{ext}} - E_{ext}$ phase space from Device B2 (hBN/Gr/$ReS_2$/hBN), demonstrating a similar variation in the position of the CNP.

To investigate the origin of this phenomenon, we study the $n_{\text{ext}} - E_{ext}$ phase spaces of $R_{\text{S}}$ from Devices B1 and B2 (featuring a moiré at the top hBN/SLG interface) while varying $E_{ext}$ in both ascending and descending directions.[41] To execute this, we first keep the $E_{ext}$ fixed to a certain value and sweep $n_{\mathrm{ext}}$ from -0.1 to 0.7 x 10$^{12}$ cm$^{-2}$. Then, we sweep back to $n_{\mathrm{ext}} = -0.1\mathrm{x}10^{12}$ cm$^{-2}$ without collecting data. After that, the electric field is varied by a small step, and the $n_{\mathrm{ext}}$ sweep is repeated. **Figure 2**a and b present $n_{\mathrm{ext}} - E_{ext}$ phase spaces for Device B1, where $E_{ext}$ is varied in ascending and descending directions, respectively. The resistance of different sweeping directions is denoted as $R_S^+$ for the ascending sweep and $R_S^-$ for the descending sweep. A comparison between **Figure 2**a and b reveals a subtle difference in the position of the CNP for ascending and descending $E_{ext}$ sweep directions, indicating a $E_{ext}$-dependent hysteresis.

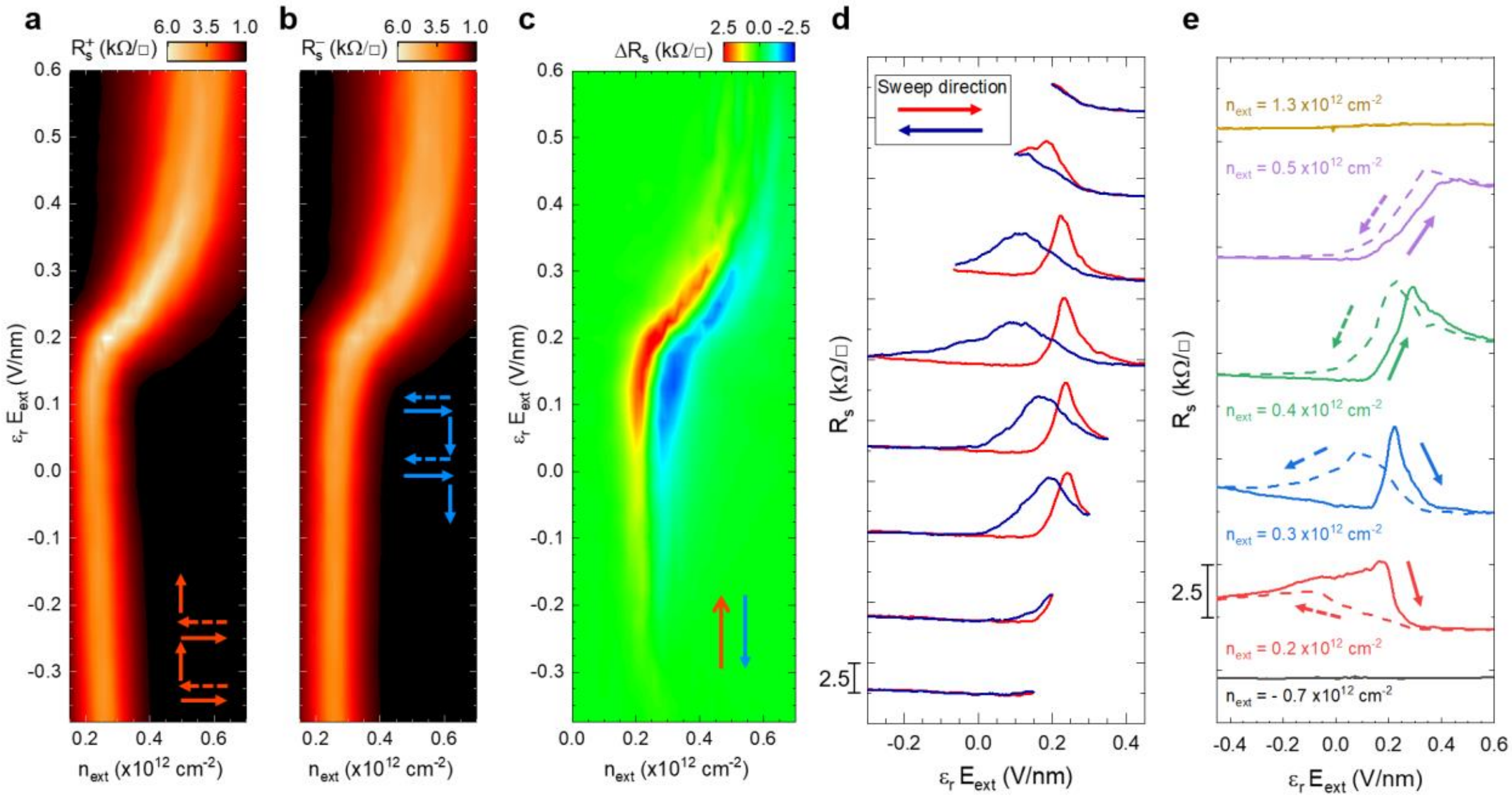

**Figure 2.** **(a, b)**, $n_{\mathrm{ext}} - E_{ext}$ phase diagram of Device B1 for forward (backward) $E_{ext}$ sweep directions. The data is collected by sweeping $n_{\mathrm{ext}}$ from the hole side ($n_{\mathrm{ext}}$<0) to the electron side ($n_{\mathrm{ext}}$>0). Then, we sweep $n_{\mathrm{ext}}$ back to starting value (dashed arrow). After that, $E_{ext}$ is increased **(a)** or decreased **(b)** and the $n_{\mathrm{ext}}$ sweep cycle is repeated. $R_S^+$ denotes the sheet resistance for the forward $E_{ext}$ sweep and $R_s^-$ for the backward sweep. **(c)** Difference between the two sweep cycles $\Delta R_{\mathrm{s}} = R_S^+ - R_S^-$. **(d)** Minor hysteretic loops at fixed $n_{\mathrm{ext}} = 0.3$ x10$^{12}$ cm$^{-2}$ for different sweep ranges. **(e)** $R_{\mathrm{s}}$ versus $E_{ext}$ for different values of the carrier concentration $n_{\mathrm{ext}}$. The solid lines represent forward sweeps from $-E_{ext}$ to $+E_{ext}$. The dashed lines are the backward sweeps from $+E_{ext}$ to $-E_{ext}$.

To further elucidate the $E_{ext}$ dependent hysteresis, the change in $R_{\mathrm{S}}$ ($\Delta R_{\mathrm{S}} = R_S^+ - R_S^-$) between ascending and descending $E_{ext}$ sweeps is presented in **Figure 2**c. At large positive and negative electric field ($\varepsilon_r E_{ext} < -0.2$ V/nm and $\varepsilon_r E_{ext} > 0.4$ V/nm), the sheet resistance $R_{\mathrm{S}}$ converges for both sweep directions, resulting in $\Delta R_S \approx 0$. However, at intermediate values of $E_{ext}$, a prominent hysteresis is observed, resulting in a large $\Delta R_S$. To further visualize this ferroelectric behaviour, we study the hysteresis for different $E_{ext}$ sweep ranges at $n_{ext}$= 0.3 x$10^{12}$ cm$^{-2}$ (**Figure 2**d). When the sweep range remains within $\varepsilon_r E_{ext}$ < 0.15 V/nm (bottom-most loop in **Figure 2**d), the trace and retrace scans fall on top of each other, indicating the absence of any hysteresis. As the range is increased to $\varepsilon_r E_{ext} \approx 0.2$ V/nm, a small minor loop appears, indicating a partial switching of the polarization. With further increase in the electric field range, full switching occurs. An opposite minor loop occurs for $0.15 < \varepsilon_r E_{ext} < 0.45$ V/nm, whereas no hysteretic behaviour is observed for $\varepsilon_r E_{ext} > 0.2$ V/nm. To further investigate this hysteretic behaviour, we sweep $D$ at several fixed values of $n_{ext}$. $R_S$ vs. $E_{ext}$ loops for a few fixed $n_{ext}$ are shown in **Figure 2**e. All the hysteresis curves follow a counterclockwise direction, resembling conventional ferroelectric behaviour. When fixing $n_{ext}$ close to the CNP ($n_{ext} = 0.3$ x$10^{12}$ cm$^{-2}$), the magnitude of the hysteresis is strongest with two distinct switching peaks at $\varepsilon_r E_{ext}$ = 0.23±0.01 V/nm and $\varepsilon_r E_{ext}$ = 0.11±0.02 V/nm. Importantly, control measurements performed at varying sweep delays (SI section 6) show that the CNP position remains stable, with the same well-defined switching fields, thereby excluding conventional charge-trapping mechanisms that would strongly depend on sweep rate. While changing the carrier concentration to $n_{ext} = 0.4$ x$10^{12}$ cm$^{-2}$ and further increasing it to $n_{ext} =$ 0.5 x$10^{12}$ cm$^{-2}$, the hysteresis peaks shift toward higher $E_{ext}$. The opposite trend is observed on the hole conduction side at $n_{ext} =$ 0.2 x$10^{12}$ cm$^{-2}$. At high carrier concentrations $n_{ext} = 1.3$ x$10^{12}$ cm$^{-2}$ and $n_{ext} = -0.7$ x$10^{12}$ cm$^{-2}$, no hysteretic behaviour occurs. We note that the hysteretic behaviour is more pronounced near the deviation of the CNP in the $n_{ext} - E_{ext}$ phase space. This indicates that the interfacial polarization charges are increasingly screened by injected carriers in the channel. A similar trend in the electronic $n_{ext} - E_{ext}$ phase space is also observed in Device B2, hBN/SLG/$ReS_2$/hBN (details in SI section 5). This strongly suggests that the effect originates from the top hBN/SLG moiré superlattice, regardless of the specific bottom interfaces.

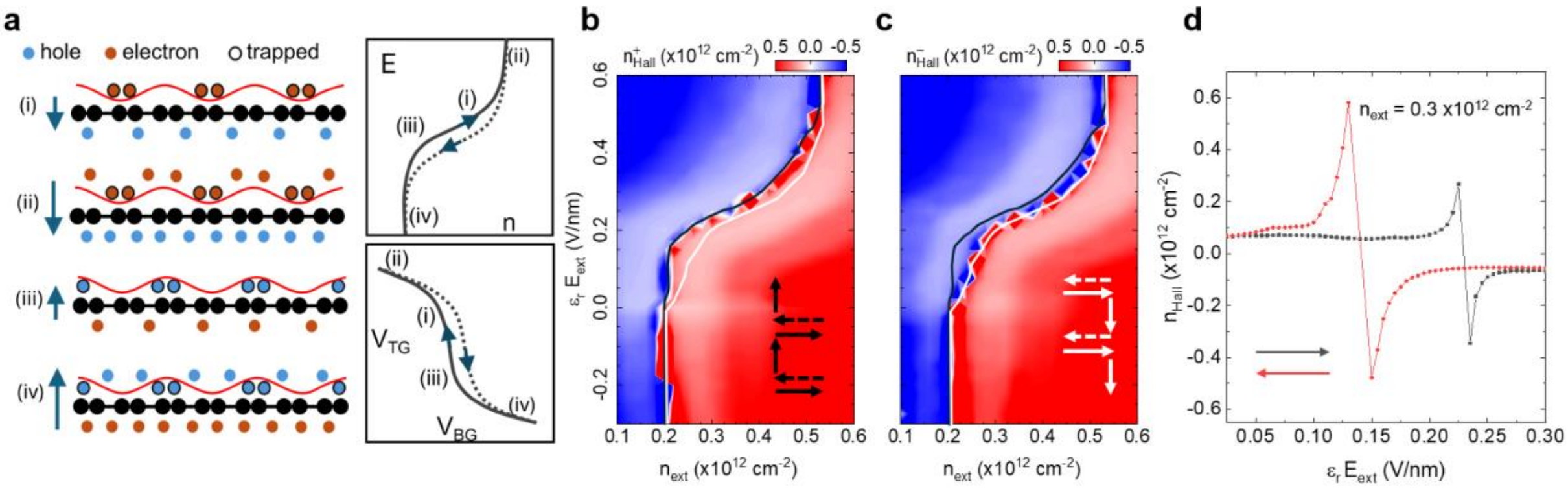

**Figure 3. (a)** Schematic illustration of the charge-trapping mechanism at the top moiré interface. Black, brown and blue spheres represent carbon atoms, electrons and holes, respectively; spheres with black outlines indicate trapped charges. The red trace represents the moiré potential at the top interface. Processes (i) and (iii) correspond to electron and hole trapping due to the moiré potential at the top interface. Processes (ii) and (iv) represent a saturation region where all trap sites are filled. Right: Evolution of the CNP in $n_{\text{ext}} - E_{ext}$ phase space (**top right**) and dual gated configuration (**bottom right**). Solid and dashed lines indicate trace and retrace sweep directions, respectively. **(b, c)** $n_{\text{ext}} - E_{ext}$ phase space of $n_{\text{Hall}}^{+}$ for forward sweep **(b)** and $n_{\text{Hall}}^{-}$ for backward sweep **(c)** extracted from Hall measurements. The arrows indicate the sweep direction of each phase space. For clarity, we also indicate the evolution of the CNP in white and black for forward and backward $E_{ext}$ sweep respectively. **(d)** Hall carrier concentration $n_{\text{Hall}}$ vs $E_{ext}$ sweep at fixed $n_{\text{ext}} = 0.3$ x$10^{12}$ cm$^{-2}$.

We attribute this phenomenon to interfacial charge trapping and detrapping mediated by the moiré potential, similar to doubly aligned SLG. **Figure 3**a schematically illustrates this behaviour. The application of $E_{ext}$ at fixed $n_{\text{ext}}$ can be interpreted as the combined effect of the top and bottom gates accumulating opposite charges on the top and bottom surfaces of graphene. When $E_{ext}$ is increased from zero to a small positive value (processes i and iii), both gates induce equal densities of electrons and holes in the channel. However, due to the presence of the moiré potential, carriers injected by the top gate—electrons in process i and holes in process iv—become trapped. As a result, only carriers accumulated from the bottom gate contribute to the current, leading to partial screening and bending of the CNP trajectory. With further increase in $E_{ext}$, once the trap sites are fully saturated, both gates induce equal numbers of mobile electrons and holes (processes ii and iv), diminishing the screening effect and stabilizing the carrier density. This dynamic trapping and subsequent release of carriers underlies the observed hysteresis in the system. It is worth noting that the experimentally observed deviation of the CNP is diagonally shifted from $E_{ext} = 0, n_{\text{ext}} = 0$ due to the presence of additional charge disorder independent of moiré traps commonly observed in graphene FETs. The trapped charges induced by the top gate, combined with the free charges induced by the bottom gate, lead to an interfacial polarization ($P_{in}$), the sign of which can be reversed by $E_{ext}$.

To quantify the polarization in the system, we directly measure the carrier concentration $n_{\mathrm{Hall}}$ in our device using Hall measurements (see SI section 2). $n_{\mathrm{ext}} - E_{ext}$ phase space $n_{\mathrm{Hall}}$ for forward and backward sweeps are presented in **Figure 3**b and c, respectively. The hole ($n_{\mathrm{Hall}} < 0$) and the electron ($n_{\mathrm{Hall}} > 0$) doped regimes are shown in blue and red, respectively. The evolution of the CNP is indicated by the black trace. The positions of the CNP for $n_{\mathrm{Hall}} = 0$ match the results obtained by longitudinal resistance in **Figure 2**a, b. To compensate for the small residual doping in device B1 and thereby probe the behavior near the CNP, we fixed the carrier concentration at $n_{\mathrm{ext}}$= 0.3 x10$^{12}$ cm$^{-2}$, as shown in **Figure 3**d. While changing $E_{ext}$, a shift in $n_{\mathrm{Hall}}$ is observed. Specifically, the carrier type indicated by $n_{\mathrm{Hall}}$ changes sign at $\varepsilon_r E_{ext}$= 0.23V/nm and $\varepsilon_r E_{ext}$= 0.13 V/nm. These values of $\varepsilon_r E_{ext}$ correspond to the position of the peaks observed in the $R_s$ $vs.$ $E_{ext}$, confirming the CNP as their origin. In the forward sweep, the injected carrier density $n_{\mathrm{ext}}$ (= 0.3 x 10$^{12}$ cm$^{-2}$) is above the CNP (= 0.2 x 10$^{12}$ cm$^{-2}$) until $\varepsilon_r E_{ext}$= 0.23 V/nm, leading to $n_{\mathrm{Hall}}$> 0, i.e. net electron conduction. At the hysteretic point, the CNP shifts to 0.5 x10$^{12}$ cm$^{-2}$, higher than the injected carrier density $n_{\mathrm{ext}}$, leading to $n_{\mathrm{Hall}}$< 0, i.e. net hole conduction. On the reverse sweep, the opposite is true, with the CNP reverting to 0.2 x 10$^{12}$ cm$^{-2}$ at a electric field of $\varepsilon_r E_{ext}$= 0.1 V/nm. Taking the difference between the two different sweep directions, the additional charge injected or withdrawn in SLG is $2\Delta n_{\mathrm{Hall}} = n_{\mathrm{Hall}}^{+} - n_{\mathrm{Hall}}^{-}$ = (0.14 ± 0.01) x10$^{12}$ cm$^{-2}$. The corresponding spontaneous polarization, compared to conventional 3D ferroelectrics, is given by $P_{3D} = e\Delta n_{\mathrm{Hall}}$ =(1.12 ± 0.08) x10$^{-2}$ μC cm$^{-2}$. To account for the dipolar contribution from both the top and bottom interfaces of the SLG, we compute the effective 2D polarization $P_{2D} = e\Delta n_{\mathrm{Hall}} d_{\mathrm{dipole}}$ = (2.91 ± 0.21) x10$^{-2}$ pC m$^{-1}$, where $d_{\mathrm{dipole}} = 0.26$ nm is the thickness of SLG.[17,31] Furthermore, from the number of additional charge injected/withdrawn due to the alignment of hBN and SLG, we can calculate the moire potential ~ 31 meV (details in SI 9), in agreement with previous theoretical prediction.[42]

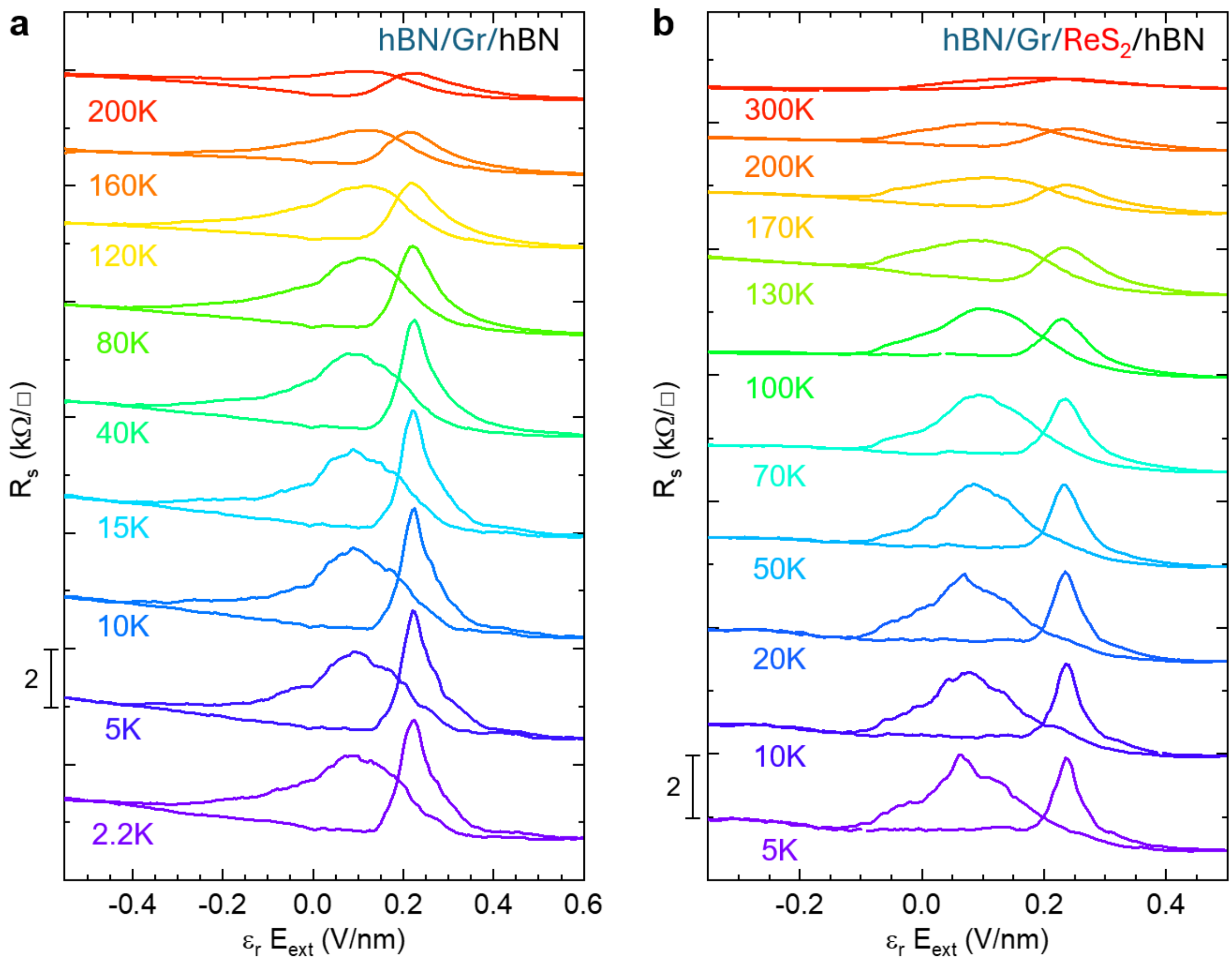

**Figure 4.** Temperature dependence of ferroelectric hysteresis in singly aligned hBN/SLG on **(a)** misaligned bottom hBN and **(b)** $ReS_2$. The traces are vertically shifted for clarity.

We now explore the temperature dependence of the observed effect. We fixed $n_{\text{ext}} = 0.3 \times 10^{12}$ cm$^{-2}$ and recorded the resistance while sweeping the electric field $E_{ext}$ at different temperatures. The results are shown in **Figure 4**. As mentioned above, the origin of the hysteresis peaks is the abnormal change in the carrier concentration in the device. Graphene's mobility decreases with increasing temperatures (see SI 8), leading to a broadening of the CNP and to an observed decrease in the magnitude of the hysteresis peaks at high temperatures.

Notably, the hysteresis switching peak positions remain nearly unchanged at $\varepsilon_r E_{ext}$= 0.1 V/nm and $\varepsilon_r E_{ext}$ = 0.23 V/nm throughout the measured temperature range. This further confirms the absence of thermally activated charge traps in our system, which would lead to a shift in the peak positions with temperature.[43,44] Rather, the hysteresis is robust and related to a temperature-invariant potential at one interface, similar to previous effects observed in the graphene/hBN system.

## 3. Conclusion

In summary, we demonstrate that robust ferroelectric hysteresis can be achieved in singly aligned hBN/SLG heterostructures, establishing that a single moiré interface is sufficient to induce ferroelectricity. Polarization switching occurs at displacement fields of ~0.1–0.23 V/nm, comparable with previous report on BLG and remains stable across the full temperature range from 2 K to 300 K, confirming excellent retention and thermal robustness. By aligning only the top interface while misaligning the bottom, and further suppressing bottom moiré formation using a $ReS_2$ spacer, we isolate the moiré-induced effects with high precision and calculated the moiré potential ~31 meV in our case. This architecture simplifies device fabrication, as it requires alignment at only one interface, and substantially reduces tolerances compared to double alignment. These findings position singly aligned SLG/hBN heterostructures as a simpler yet equally effective platform for moiré-engineered ferroelectricity, opening new opportunities for scalable, ultrathin non-volatile memory and logic devices.

## 4. Methods

Hexagonal boron nitride (hBN) is used as the top gate dielectric, and 300 nm $SiO_2$ is employed as the bottom gate dielectric (also see SI section 1). Electrical measurements were conducted using a Keithley 6221 current source synchronized with lock-in amplifiers at a reference frequency of 17.77 Hz. The input impedances of all instruments were significantly higher than the sheet resistance ($R_s$). All the measurement was conducted in a 4-probe measurement configuration to explicit avoid the contact induced hysteresis behavior. Gate voltages were applied using dual-channel DC source meters. The leakage current was less than 1nA for all the gate measurements. All experiments were carried out in a closed-cycle cryostat at 2 K unless otherwise specified.

The $n_{\mathrm{ext}} - E_{ext}$ phase space was collected by setting the external carrier concentration $n_{\mathrm{ext}} = \frac{\varepsilon_0}{e}(\frac{\epsilon_r V_{BG}}{d_{BG}} + \frac{\epsilon_r V_{TG}}{d_{TG}})$ and the vertical electric field $E_{ext}$ (the direction pointing downward from top to bottom gate) as $\varepsilon_r E_{ext} = \frac{1}{2}(\frac{V_{TG}}{d_{TG}} - \frac{V_{BG}}{d_{BG}})$, where $d_{TG}$ and $d_{BG}$ are thickness of top and bottom gate dielectric; $\epsilon_r$ is gate dielectric constant .[33,38]

To conduct Hall measurement, we applied out of plane magnetic field in two opposite directions $B = \pm 1\mathrm{T}$ and taken transverse resistance $R_{xy}^{+}$ and $R_{\mathrm{xy}}^{-}$ for forward and backward sweep direction respectively. The Hall resistance is taken by antisymmetrize transverse resistance $R_{\mathrm{Hall}}^{\pm} = (R_{xy}^{\pm(\mathrm{B=1T})} - R_{xy}^{\pm(\mathrm{B=-1T})})/2$ . From that, Hall carrier concentration is calculated by using the formula $n_{Hall}^{\pm} = -\frac{|B|}{eR_{Hall}^{\pm}}$ .

**Acknowledgements**

The authors acknowledge financial support from MICIU/AEI/10.13039/501100011033 (Grant CEX2020-001038-M), from MICIU/AEI and ERDF/EU (Projects PID2021-122511OB-I00 and PID2021-128004NB-C21), from MICIU/AEI and European Union NextGenerationEU/PRTR (Grant PCI2021-122038-2A.). Authors also acknowledge funding from the European Union's Horizon 2020 research and innovation programme under Marie Skłodowska-Curie grant agreement No. 955671, as well as from the European Union's Horizon Europe program under grant agreement No. 101046231. B.Q.T acknowledges funding from MICIU/AEI and ESF+ (Fellowship PRE2022-103674). T.A. and S.L. acknowledge funding from the European Union (Grant Agreements 101107842 and 101151842, respectively). K.W. and T.T. acknowledge support from the JSPS KAKENHI (Grant Numbers 21H05233 and 23H02052) and World Premier International Research Center Initiative (WPI), MEXT, Japan. M.G. acknowledges funding from MICIU/AEI and European Union NextGenerationEU/PRTR (Fellowship RYC2021-031705-I).

**References**

[1] J. F. Scott, C. A. Paz De Araujo, *Science* **1989**, *246*, 1400.
[2] A. Q. Jiang, C. Wang, K. J. Jin, X. B. Liu, J. F. Scott, C. S. Hwang, T. A. Tang, H. B. Lu, G. Z. Yang, *Advanced Materials* **2011**, *23*, 1277.
[3] S. Mathews, R. Ramesh, T. Venkatesan, J. Benedetto, *Science* **1997**, *276*, 238.
[4] R. C. G. Naber, C. Tanase, P. W. M. Blom, G. H. Gelinck, A. W. Marsman, F. J. Touwslager, S. Setayesh, D. M. De Leeuw, *Nature Mater* **2005**, *4*, 243.
[5] Z. Wen, C. Li, D. Wu, A. Li, N. Ming, *Nature Mater* **2013**, *12*, 617.
[6] J. F. Scott, *Science* **2007**, *315*, 954.
[7] G. Catalan, J. F. Scott, *Advanced Materials* **2009**, *21*, 2463.
[8] J. Valasek, *Phys. Rev.* **1921**, *17*, 475.
[9] G. Busch, P. Scherrer, *Ferroelectrics* **1987**, *71*, 15.
[10] A. Von Hippel, R. G. Breckenridge, F. G. Chesley, L. Tisza, *Ind. Eng. Chem.* **1946**, *38*, 1097.
[11] H. Takasu, *Journal of Electroceramics* **2000**, *4*, 327.
[12] X. Chen, X. Han, Q. Shen, *Adv Elect Materials* **2017**, *3*, 1600460.
[13] U. Schroeder, M. H. Park, T. Mikolajick, C. S. Hwang, *Nat Rev Mater* **2022**, *7*, 653.
[14] J. P. B. Silva, R. Alcala, U. E. Avci, N. Barrett, L. Bégon-Lours, M. Borg, S. Byun, S.-C. Chang, S.-W. Cheong, D.-H. Choe, J. Coignus, V. Deshpande, A. Dimoulas, C. Dubourdieu, I. Fina, H. Funakubo, L. Grenouillet, A. Gruverman, J. Heo, M. Hoffmann, H. A. Hsain, F.-T. Huang, C. S. Hwang, J. Íñiguez, J. L. Jones, I. V. Karpov, A. Kersch, T. Kwon, S. Lancaster, M. Lederer, Y. Lee, P. D. Lomenzo, L. W. Martin, S. Martin, S. Migita, T. Mikolajick, B. Noheda, M. H. Park, K. M. Rabe, S. Salahuddin, F. Sánchez, K. Seidel, T. Shimizu, T. Shiraishi, S. Slesazeck, A. Toriumi, H. Uchida, B. Vilquin, X. Xu, K. H. Ye, U. Schroeder, *APL Materials* **2023**, *11*, 089201.
[15] Z. Fei, W. Zhao, T. A. Palomaki, B. Sun, M. K. Miller, Z. Zhao, J. Yan, X. Xu, D. H. Cobden, *Nature* **2018**, *560*, 336.
[16] C. R. Woods, P. Ares, H. Nevison-Andrews, M. J. Holwill, R. Fabregas, F. Guinea, A. K. Geim, K. S. Novoselov, N. R. Walet, L. Fumagalli, *Nat Commun* **2021**, *12*, 347.

[17] K. Yasuda, X. Wang, K. Watanabe, T. Taniguchi, P. Jarillo-Herrero, *Science* **2021**, *372*, 1458.
[18] M. Vizner Stern, Y. Waschitz, W. Cao, I. Nevo, K. Watanabe, T. Taniguchi, E. Sela, M. Urbakh, O. Hod, M. Ben Shalom, *Science* **2021**, *372*, 1462.
[19] X. Wang, K. Yasuda, Y. Zhang, S. Liu, K. Watanabe, T. Taniguchi, J. Hone, L. Fu, P. Jarillo-Herrero, *Nat. Nanotechnol.* **2022**, *17*, 367.
[20] P. Meng, Y. Wu, R. Bian, E. Pan, B. Dong, X. Zhao, J. Chen, L. Wu, Y. Sun, Q. Fu, Q. Liu, D. Shi, Q. Zhang, Y.-W. Zhang, Z. Liu, F. Liu, *Nat Commun* **2022**, *13*, 7696.
[21] C. Wang, L. You, D. Cobden, J. Wang, *Nat. Mater.* **2023**, *22*, 542.
[22] Q. Liu, S. Cui, R. Bian, E. Pan, G. Cao, W. Li, F. Liu, *ACS Nano* **2024**, *18*, 1778.
[23] K. S. Novoselov, A. Mishchenko, A. Carvalho, A. H. Castro Neto, *Science* **2016**, *353*, aac9439.
[24] A. K. Geim, I. V. Grigorieva, *Nature* **2013**, *499*, 419.
[25] H. Yang, M. Gobbi, F. Herling, V. T. Pham, F. Calavalle, B. Martín-García, A. Fert, L. E. Hueso, F. Casanova, *Nat Electron* **2024**, *8*, 15.
[26] C. K. Safeer, J. Ingla-Aynés, F. Herling, J. H. Garcia, M. Vila, N. Ontoso, M. R. Calvo, S. Roche, L. E. Hueso, F. Casanova, *Nano Lett.* **2019**, *19*, 1074.
[27] L. J. McGilly, A. Kerelsky, N. R. Finney, K. Shapovalov, E.-M. Shih, A. Ghiotto, Y. Zeng, S. L. Moore, W. Wu, Y. Bai, K. Watanabe, T. Taniguchi, M. Stengel, L. Zhou, J. Hone, X. Zhu, D. N. Basov, C. Dean, C. E. Dreyer, A. N. Pasupathy, *Nat. Nanotechnol.* **2020**, *15*, 580.
[28] H. Yang, B. Martín-García, J. Kimák, E. Schmoranzerová, E. Dolan, Z. Chi, M. Gobbi, P. Němec, L. E. Hueso, F. Casanova, *Nat. Mater.* **2024**, *23*, 1502.
[29] E. Y. Andrei, D. K. Efetov, P. Jarillo-Herrero, A. H. MacDonald, K. F. Mak, T. Senthil, E. Tutuc, A. Yazdani, A. F. Young, *Nat Rev Mater* **2021**, *6*, 201.
[30] B. Hunt, J. D. Sanchez-Yamagishi, A. F. Young, M. Yankowitz, B. J. LeRoy, K. Watanabe, T. Taniguchi, P. Moon, M. Koshino, P. Jarillo-Herrero, R. C. Ashoori, *Science* **2013**, *340*, 1427.
[31] L. Zhang, J. Ding, H. Xiang, N. Liu, W. Zhou, L. Wu, N. Xin, K. Watanabe, T. Taniguchi, S. Xu, *Nat Commun* **2024**, *15*, 10905.
[32] R. Niu, Z. Li, X. Han, Z. Qu, D. Ding, Z. Wang, Q. Liu, T. Liu, C. Han, K. Watanabe, T. Taniguchi, M. Wu, Q. Ren, X. Wang, J. Hong, J. Mao, Z. Han, K. Liu, Z. Gan, J. Lu, *Nat Commun* **2022**, *13*, 6241.
[33] Z. Zheng, Q. Ma, Z. Bi, S. de la Barrera, M.-H. Liu, N. Mao, Y. Zhang, N. Kiper, K. Watanabe, T. Taniguchi, J. Kong, W. A. Tisdale, R. Ashoori, N. Gedik, L. Fu, S.-Y. Xu, P. Jarillo-Herrero, *Nature* **2020**, *588*, 71.
[34] Z. Zhu, S. Carr, Q. Ma, E. Kaxiras, *Phys. Rev. B* **2022**, *106*, 205134.
[35] K. Roy, T. Ahmed, H. Dubey, T. P. Sai, R. Kashid, S. Maliakal, K. Hsieh, S. Shamim, A. Ghosh, *Advanced Materials* **2018**, *30*, 1704412.
[36] S. Zhang, N. Mao, N. Zhang, J. Wu, L. Tong, J. Zhang, *ACS Nano* **2017**, *11*, 10366.
[37] M. Ramos, T. Ahmed, B. Q. Tu, E. Chatzikyriakou, L. Olano-Vegas, B. Martín-García, M. R. Calvo, S. S. Tsirkin, I. Souza, F. Casanova, F. De Juan, M. Gobbi, L. E. Hueso, *Nano Lett.* **2024**, *24*, 14728.
[38] T. Ahmed, H. Varshney, B. Q. Tu, K. Watanabe, T. Taniguchi, M. Gobbi, F. Casanova, A. Agarwal, L. E. Hueso, *Small* **n.d.**, *n/a*, 2501426.
[39] M. Yankowitz, J. Xue, D. Cormode, J. D. Sanchez-Yamagishi, K. Watanabe, T. Taniguchi, P. Jarillo-Herrero, P. Jacquod, B. J. LeRoy, *Nature Phys* **2012**, *8*, 382.
[40] R. Ribeiro-Palau, C. Zhang, K. Watanabe, T. Taniguchi, J. Hone, C. R. Dean, *Science* **2018**, *361*, 690.
[41] T. Ahmed, K. Watanabe, T. Taniguchi, F. Casanova, L. E. Hueso, *Advanced Materials* **2025**, e09837.

[42] J. Jung, A. M. DaSilva, A. H. MacDonald, S. Adam, *Nat Commun* **2015**, *6*, 6308.
[43] M. H. Yusuf, B. Nielsen, M. Dawber, X. Du, *Nano Lett.* **2014**, *14*, 5437.
[44] T. Feng, D. Xie, G. Li, J. Xu, H. Zhao, T. Ren, H. Zhu, *Carbon* **2014**, *78*, 250.

**Supporting Information**

The Supporting information (SI) is available. It contains the following sections. Section 1: Fabrication method. Section 2: Raman and quantum Hall characteristics of SLG. Section 3: Twisted angle and moiré wavelength calculation. Section 4: $V_{TG}$-$V_{BG}$ phase space of device B1. Section 5: Hysteretic behavior in aligned Gr/hBN on $ReS_2$. Section6: Stabilities of hysteresis. Section 7: $ReS_2$ transport properties. Section 8: SLG mobilities at different temperatures. Section 9: Moire potential calculation. Section 10:Threshold and peak values of hysteresis.